\documentclass[preprint]{aastex}
\usepackage{graphicx}
\usepackage{amsmath,amssymb}
\usepackage{float}
\title{On the origin of inner coma structures observed by Rosetta during a diurnal rotation of comet 67P/Churyumov-Gerasimenko.}
\author{Tobias~Kramer$^{a,b}$, Matthias~Noack$^{a}$\\
{\small $^a$Konrad-Zuse-Zentrum f\"ur Informationstechnik,  Takustr.~7, 14195 Berlin, Germany}\\
{\small $^b$Department of Physics, Harvard University, 12 Oxford St, Cambridge, MA 02138, U.S.A.}
}
\date{Published in \textit{The Astrophysical Journal Letters}, \textbf{823}, L11 (2016)}

\begin{document}

\begin{abstract}
The Rosetta probe around comet 67P/Churyumov-Gerasimenko (67P) reveals an anisotropic dust distribution of the inner coma with jet-like structures.
The physical processes leading to jet formation are under debate, with most models for cometary activity focusing on localised emission sources, such as cliffs or terraced regions.
Here we suggest, by correlating high-resolution simulations of the dust environment around 67P with observations, that the anisotropy and the background dust density of 67P originate from dust released across the entire sunlit surface of the nucleus rather than from few isolated sources.
We trace back trajectories from coma regions with high local dust density in space to the non-spherical nucleus and identify two mechanisms of jet formation:
areas with local concavity in either two dimensions or only one. 
Pits and craters are examples of the first case, the neck region of the bilobed nucleus of 67P for the latter one.
The conjunction of multiple sources in addition to dust released from all other sunlit areas results in a high correlation coefficient ($\sim$0.8) of the predictions with observations during a complete diurnal rotation period of 67P.
\end{abstract}

\maketitle

\section{Introduction}

Solar illumination drives the gas and dust emission from comets by sublimating the ice of the frozen dust-gas conglomerate.
Embedded dust particles are accelerated by the expanding gas in the outer porous mantle of the comet and form the innermost coma around the nucleus (\cite{Huebner2006}).
Starting with the 1P/Halley flyby, regions with increased dust intensity within several cometary radii have been imaged and connected to active and inactive surface areas (\cite{Whipple1982,Keller1994,Combi2012,BruckSyal2013,Belton2013}), or alternatively to the general shape of the nucleus (\cite{Crifo2002,Zakharov2009}).
Specific surface features linked to jet formation include active pits (\cite{Vincent2015}), terraced regions (\cite{Farnham2013}), and cliffs (\cite{Vincent2015a}).
In addition cometary {dust and gas emission} highly varies according to the rapidly changing temperature conditions driven by solar illumination (\cite{Ali-Lagoa2015a,Lara2015}).
So far, the complex {dust} activity with temporally and spatially varying source patterns has precluded a detailed prediction of the dust distribution observed in the innermost coma of 67P besides for two specific cases (\cite{Kramer2015a,Marschall2015}). 
The conclusions drawn from these studies are limited by the lack of testing the theoretical prediction across a complete diurnal rotation of the nucleus.
One obstacle for a predictive model has been the computational complexity to describe the cometary gas and dust in three dimensions emitted from a rotating cometary nucleus.
Here, we report results  for a high-resolution three-dimensional model of the dust environment around 67P predicted across the entire rotation period of the nucleus, which besides the time-dependent gas-dust interaction, also takes into account the rotation and detailed shape of the nucleus  (\cite{Kramer2015,Kramer2015a}).

\section{Cometary shape and emission condition}

The long-term mission of Rosetta following 67P (\cite{Schulz2009}) provides detailed shape models of the nucleus and records image sequences (\cite{Sierks2015}) to test cometary {dust} models (\cite{Vincent2015a}).
To establish an unambiguous link between photographed intensities and sources of cometary activity requires a model for the flow of dust particles in combination with the assignment of dust sources on the surface.
The conceptually simplest models of cometary activity describes the emission of gas (and dust) from a spherical nucleus with homogeneous surface activity (\cite{Haser1957,Combi2004}).
Explaining jet-like structures arising from such a spherical nucleus requires to restrict the surface gas and dust emission to a small number of active areas (\cite{Combi2012}).
The bilobed shape and complex surface morphology of several nuclei targeted by flyby missions suggests an alternative scenario for dust emission and the origin of a structured innermost coma, including highly concentrated jets.
We maintain the earliest assumption of homogeneous surface activity across the entire nucleus, but pay full attention to the non-spherical shape and rotation of the nucleus, including the surface orography.
The model could be augmented at a later stage to incorporate different activity zones across the surface.
However, it is of interest to find out to what extent already the homogeneous {gas and dust} activity across the entire nucleus describes the Rosetta data, since a good agreement restricts further parameters required to describe additional effects such as varying surface composition.
To test this hypothesis, we derive a uniformly remeshed high-resolution  shape model of 67P, covering the surface with 199,982 triangles of areas (248$\pm4$9)~m$^2$, from the 67P nucleus shape model given by \cite{MalmerShapeModelNovmeber2015}.
The uniformly remeshed shape obliterates the need to perform a random Monte-Carlo sampling of the velocity/position phase space around the nucleus.
Gas and dust are emitted perpendicular to the surface normal across the entire nucleus.
Cometary surfaces display complex surface textures where fractures might act as nozzles for the expanding gas in the outer crust (\cite{Gundlach2015}).
The higher gas pressure in the cracks and fractures leads to the acceleration of dust within the outer mantle. 
The observational implications of an acceleration of dust already in the outer mantle layer, as opposed to dust lifted off with zero velocity from the surface, will be discussed later.
We solve the Haser model for the gas dynamics (\cite{Haser1957,Combi2004}), adapted to the complex shape of the nucleus and for a gas composed of water molecules.
The Haser model assumes a constant and uniform gas flux per surface area across the whole cometary surface {(area $A$)}, directed along the surface normal in the body fixed frame. 
Collisions between gas molecules and the production of daughter species through ionization are neglected. 
{With the gas number density $N_{\text{gas}}^0$ at the surface, molecular mass $m_\text{gas}$ and mean gas velocity $v_{\text gas}$ directed perpendicular to the surface, the overall gas production rate
\begin{equation}
Q_\text{gas}=\int_{\mathbf{r}\in\text{surface}}  m_\text{gas}  N_\text{gas}(\mathbf{r}) \vec{v}_\text{gas}(\mathbf{r}) \!\cdot \!{\rm d} \vec{S}
\end{equation}
simplifies to $Q_\text{gas}^0=m_\text{gas} v_\text{gas} N_{\rm gas}^0 A$.
The gas moves outside the nucleus under the influence of gravitation and rotational forces with the cometary rotation axis given by \cite{Sierks2015}.
The gas velocity is set to the mean thermal velocity at temperature $T=210$~K 
$\tilde{v}_{\rm th}=\sqrt{\frac{3 k_B T}{m_{\rm gas}}}=540$~m/s  (\cite{Vincent2015a}) .
Depending on the gas density at the surface $N_{\rm gas}^0=10^{16}$~m$^{-3}~\text{--}~10^{18}$~m$^{-3}$, a total gas production rate of $Q_{\rm gas}^0=8$~kg/s$~\text{--}~800$~kg/s results.
The surface flux of the gas is assumed to be constant across the surface and in time.
The reduced gas and dust emission at the night hemisphere is taken into account at a later stage by restricting the dust emission to the illuminated parts of the surface. 
}

\section{Dust emission from the mantle}

Once the gas atmosphere has been established, dust trajectories are computed by immersing dust particles in the gas field.
The microscopic initial conditions surrounding the dust ejection into space from the mantle have a large impact on the global dust distribution around the nucleus.
Previous models of 67P assigned a zero initial velocity component to the dust at the outer surface of a non-rotating comet model (\cite{Combi2012,Marschall2015}). 
The dust acceleration is then purely driven by gas drag outside the nucleus.
Upon inclusion of the rotation of the nucleus, the zero-velocity condition results in a slow increase of the velocity and a large sidewards drift component caused by the Coriolis effect (\cite{Kramer2015a,Kramer2015}).
In a porous mantle layer dust particles are likely already accelerated inside the outer mantle layer due to the increased microscopic gas-dust interaction within cracks and pores.
Then, the dust emanates from the surface with a finite velocity vector anchored in the coordinate system attached to the rotating nucleus.
In this case and in marked contrast to the zero-velocity dust lift-off scenario, the dust velocity-vector carries along the direction of the local surface normal into space (Fig.~1).
Neglecting the rotation of the nucleus diminishes the agreement with observations, as the Coriolis effect is inversely proportional to the velocity (Fig.~4 in \cite{Kramer2015a}). 
The acceleration of a dust particle due to momentum transfer from the faster gas molecules is given in the nucleus-attached frame by
\begin{eqnarray*}
\vec{a}_{{\rm dust}}(\vec{r})
&=&\vec{a}_{\rm gas~drag}+\vec{a}_{\rm grav}+\vec{a}_{\rm centrifugal}+\vec{a}_{\rm Coriolis}\\\nonumber
&=&\frac{1}{2}C_d \alpha N_{\rm gas}  (\vec{r}) m_{\rm gas} (\vec{v}_{\rm gas}-\vec{v}_{\rm dust})|\vec{v}_{\rm gas}-\vec{v}_{\rm dust}|
\\&&
\quad\quad\quad- \nabla \phi(\vec{r})-\vec{\omega}\times(\vec{\omega}\times\vec{r})-2\vec{\omega}\times \vec{v}_{\rm dust},
\end{eqnarray*}
which includes, besides the gas drag the gravitational force, the effect of rotation (centrifugal and Coriolis forces).
The gravitational potential of the nucleus is obtained by using the expressions given by \cite{Conway2014} taking as density $470$~kg/m$^3$ (\cite{Sierks2015}).
For the standard value $C_d=2$ (\cite{Keller1994}) and the thermal gas velocity
$\tilde{v}_{\rm th}$ we obtain an overall factor
\begin{eqnarray}
c_{\rm gas~drag}
&=&3 k_B T N_{\rm gas}(\vec{r}) \; \frac{\pi R_{\rm dust}^2}{m_{\rm dust}}\\\nonumber
&=&\frac{9}{4} \frac{k_B T}{\rho_{\rm dust}}    \; \frac{N_{\rm gas}(\vec{r})}{R_{\rm dust}},
\end{eqnarray}
for the gas-dust interaction
\begin{equation}
\vec{a}_{\rm gas~drag}=c_{\rm gas~drag}\frac{(\vec{v}_{\rm gas}-\vec{v}_{\rm dust})|\vec{v}_{\rm gas}-\vec{v}_{\rm dust}|}{{|\vec{v}_{\rm gas}|}^2},
\end{equation}
where $R_{\rm dust}$ denotes the radius of the dust particle and  $m_{\rm dust}=\frac{4\pi}{3}R_{\rm dust}^3\rho_{\rm dust}$ its mass.
Only the ratio of $N_{\rm gas}:R_{\rm dust}$ determines the gas-drag contribution.
{For 67P, grain sizes from 0.1--1~mm have been reported to dominate the coma brightness \cite{Fulle2015}.
All shown simulations correspond to a ratio of 
$N_{\rm gas}:R_{\rm dust}=10^{18}$~m$^{-3}$~:~$100$~$\mu$m with dust density $\rho_{\rm dust}=1000$~kg/m$^3$ (equivalently $10^{19}$~m$^{-3}$~:~$1000$~$\mu$m or simultaneously increased particle mass and surface gas densities).
}
To establish an initially uniform dust distribution on the surface, 199,982 dust particles are distributed on the homogeneously meshed shape model of the nucleus.
To incorporate the directionality imposed by dust coming out of cracks and pores {with increased gas pressure due to confined space}, we assume an initial value of the dust velocity of $2$~m/s directed along the surface normal, corresponding to twice the mean escape velocity from 67P without atmosphere (\cite{Sierks2015}).
Similar results are obtained for a value of $1$~m/s (\cite{Kramer2015a}).
{Outside the porous surface, the expansion of the gas leads to a diffusive mixing from adjacent surface areas and results in a less collimated gas flow compared to the dust, \cite{Kramer2015}. 
}
The dust positions and velocities are integrated using a 4th order Runge-Kutta method with a time step of 1~s. 
The dust moves at a much slower velocity compared to the gas velocity  within the innermost coma (\cite{Keller1994,Crifo2005a}).
Heavier dust particles, ejected with a slower velocity than the escape velocity, re-collide with the nucleus and are strongly affected by the Coriolis effect (\cite{Thomas2015,Kramer2015}).
Faster dust particles escape the gravitational field of the nucleus and get further accelerated by momentum transferred from the gas.
In agreement with the recorded dust speeds by the Rosetta GIADA instrument (\cite{Fulle2015,Rotundi2015}), velocities of $10$~m/s are reached within $30$~km, corresponding to 1~h of travel time.

\section{Classification of predicted dust coma structures}
The spatial dust density is computed in a 30~km volume around the comet discretized in 50~m sized cells. 
With increasing distance from the centre of the comet, the initially homogeneous dust distribution turns into a complex pattern of locally increased dust density (Fig.~2{\sf a}), referred to as primary jet structures.
Besides these isolated spots, the dashed line in Fig.~2{\sf a} indicates a fainter, but extended band of higher dust density above the neck, referred to as secondary jet structure.
To study the origin of primary jet structures, we search for regions of high dust densities within 3-6~km distance from the nucleus.
A threshold of 1/8 of the maximal dust density has been chosen to map out the 100 dominant dust concentrations in three-dimensional space.
We identify individual jets by a connected component analysis and trace them back to their surface source-areas (Fig.~2{\sf b}).
The common feature of the associated surface sources is their concave shape associated with a positive Gaussian curvature along the two principal directions, such as pits, short valleys, and larger smooth, but still concave, plains.
The importance of concave regions for jet formation is explained by the initial velocity of the gas and dust emanating from the surface.
The (averaged) momentum of gas and dust grains emitted from surface pores is directed along the surface normal and thus mirrors the local orography (Fig.~1).
The collimation due to concave surfaces is shown for the source areas of 13 primary jets in Fig.~3{\sf a,b}.
Depending on the view direction through the coma towards the nucleus, multiple primary jet regions overlay and intensify the dust column-density along the line of sight, Fig.~3{\sf e}.
The secondary jet band has a different origin connected to the concave neck region with one positive curvature and one negative curvature along the other principal axis.
The focusing effect is further amplified by an increased gas drag and dust density above the neck due to the confluence of gas and dust from both walls of the neck valley.

\section{Correlation of the predicted coma with observations}

To evaluate to what extent the hypothesis of a homogeneous surface activity matches the observed dust intensities around 67P,
we compare the predicted dust densities to Rosetta observations acquired during a diurnal rotation period on 12~April 2015.
Recorded image intensities result from accumulated sunlight scattered from dust particles along the line of sight.
{Following \cite{Fink2015} we assume an optically thin coma for 67 with a single dominating particle size and neglect multiple light scattering.}
The observation of jets in an inhomogeneous dust coma depends on the peculiarities of the viewing geometry of the observing instruments, or the point of view when evaluating simulation data.
The primary jets remain visible for a wide range of viewing angles, but their conjunctions along the line-of-sight vary. 
Secondary jet structures are only observed in case of a parallel alignment of the viewing axis with the plane of increased dust density.
For 67P, high light intensities are recorded above the concave neck region if the line of sight aligns with the v-shaped valley.
Then, the less dense, but aligned along the line-of-sight, secondary band structure (Fig.~2{\sf a}) contributes to the spatially integrated light intensity. 
Fig.~4 shows the predicted column density of the innermost coma structure at two different times in direct comparison with OSIRIS/NAC data from Fig.~A.4 by \cite{Vincent2015a}.
The observation point is chosen at $\sim 150$~km distance from 67P and the orientations are matching the Rosetta OSIRIS/NAC camera field of view.
The circled features mark the highest imaged intensities and have a one-to-one correspondence in the predicted column densities.
No additional dust activity pattern is fitted to the model, besides that the illumination, {including self-shadowing}, is taken into account.
Since about half the cometary surface is shadowed at any given time, we assume that this leads to a decrease in local surface temperature and dust emission from shadowed areas.
A twenty minutes time-lag for the onset of increased local activity after crossing the terminator has been estimated from observations and thermal models (\cite{Vincent2015a,Ali-Lagoa2015a}).
{We neglect this time-lag as it is short compared to the 12-hour rotation period. Only dust grains originating from sunlit areas contribute to the column density. Changes of the surface dust emission during the propagation time are not taken into account.
}
The conjunction of primary jets starting from the illuminated parts of the nucleus leads to the ray-type structures in the integrated column density visible in Fig.~4.

Fig.~5 extends the comparison to twelve snap shots during the 12-hour rotation period of the comet.
Restricting the surface activity only to concave areas leads to gaps in the dust density (Fig.~3{\sf c,d}), not observed by Rosetta (Fig.~4{\sf a,c} and Fig.~5{\sf a}).
The column densities vary due to the changing contributions and alignments of the primary jets and the homogeneous emission background.
To quantify the similarity between predicted dust column-density and measured light intensity, we analyse the correlation between the simulated and observed data on an annulus enclosing the nucleus (Fig.~5).
Both, the smooth background and the additional intensity modulations caused by the conjunction of multiple jets, correlate highly with the observed light intensities.
As exemplary case, we discuss the 12:12 viewing conditions, (Fig.~5{\sf a} and Fig.~4).
The increased column density originates from the alignment of the series of 13 primary jets shown in Fig.~3 and gives a high correlation coefficient (0.90) between predicted column density and observed light intensity. 
The lowest correlation coefficient (0.47) is found for Fig.~5{\sf h}, where the simplified activity model underestimates the intensity of secondary jet structures above the neck.
{
An increased neck gas flux (\cite{Ali-Lagoa2015a,Bieler2015}) might explain this case.}
The average correlation coefficient of 0.80 between the simulated dust column-density and observed light intensity over the diurnal period signifies a close match between the homogeneous activity model and observation.

\section{Conclusion}

The residual differences between a strictly homogeneous emission model and observations could be explained by a varying surface composition, local variations in outgassing rates, the accuracy of the shape model, or {light scattering across different particle sizes (\cite{Fink2015}).}
{Obtaining the absolute dust density requires to further constrain the dust particle acceleration and the gas pressure within meters from the surface by observations.}
The homogeneous activity model already reproduces the {relative dust densities}, including jet-like features, at locations observed by Rosetta in the innermost coma around 67P.
This indicates that the homogeneous model is a suitable assumption which highly correlates with the imaged coma light-intensities.
The three-dimensional dust tracing analysis shows that photographed jets are often a conjunction of multiple aligned jet structures on top of a non-collimated, homogeneously released dust background. 
Adding a locally varying activity profile on top of the homogeneous model is in principle possible, but restricting the dust emission to only surface concavities reduces the agreement with observations.
{A rapid acceleration of dust in the vicinity of the surface results in a coma mirroring the surface orography.}
This supports the hypothesis that dust emanates from fractures and subsurface pores in the mantle (\cite{Vincent2015a}).
{
For a slower lift-off velocity, the rotation of the nucleus leads to a dispersion and drift of the dust, \cite{Kramer2015a}.}
A possible implication of the underlying homogeneous emission model is that the overall surface ablation of the illuminated nucleus proceeds in a more uniform way (\cite{Cheng2013,Schulz2015}) compared to models of isolated dust emission sources.

{
The work was supported by the North-German Supercomputing Alliance (HLRN) and by the German Research Foundation within the Heisenberg program (T.K.). We thank M.~Malmer for providing the high-resolution shape model and M.~L\"auter for discussions.
We thank the OSIRIS team for providing the NAC data sets.
OSIRIS was built by a consortium led by the Max Planck Institut f\"ur Sonnensystemforschung, G\"ottingen, Germany, in collaboration with CISAS, University of Padova, Italy, the Laboratoire d'Astro\-physique de Marseille, France, the Instituto de Astrofisica de Andalucia, CSIC, Granada, Spain, the Scientific Support Office of the European Space Agency, Noordwijk, The Netherlands, the Instituto Nacional de Tecnica Aeroespacial, Madrid, Spain, the Universidad Politecnica de Madrid, Spain, the Department of Physics and Astronomy of Uppsala University, Sweden, and the Institut f\"ur Datentechnik und Kommunikationsnetze der Technischen Universit\"at Braunschweig, Germany.
The OSIRIS team acknowledges support of the national funding agencies of Germany (DLR), France (CNES), Italy (ASI), Spain (MINECO), Sweden (SNSB), and the ESA Technical Directorate.
}

\begin{figure}[h]
\begin{center}
\includegraphics[width=0.35\textwidth]{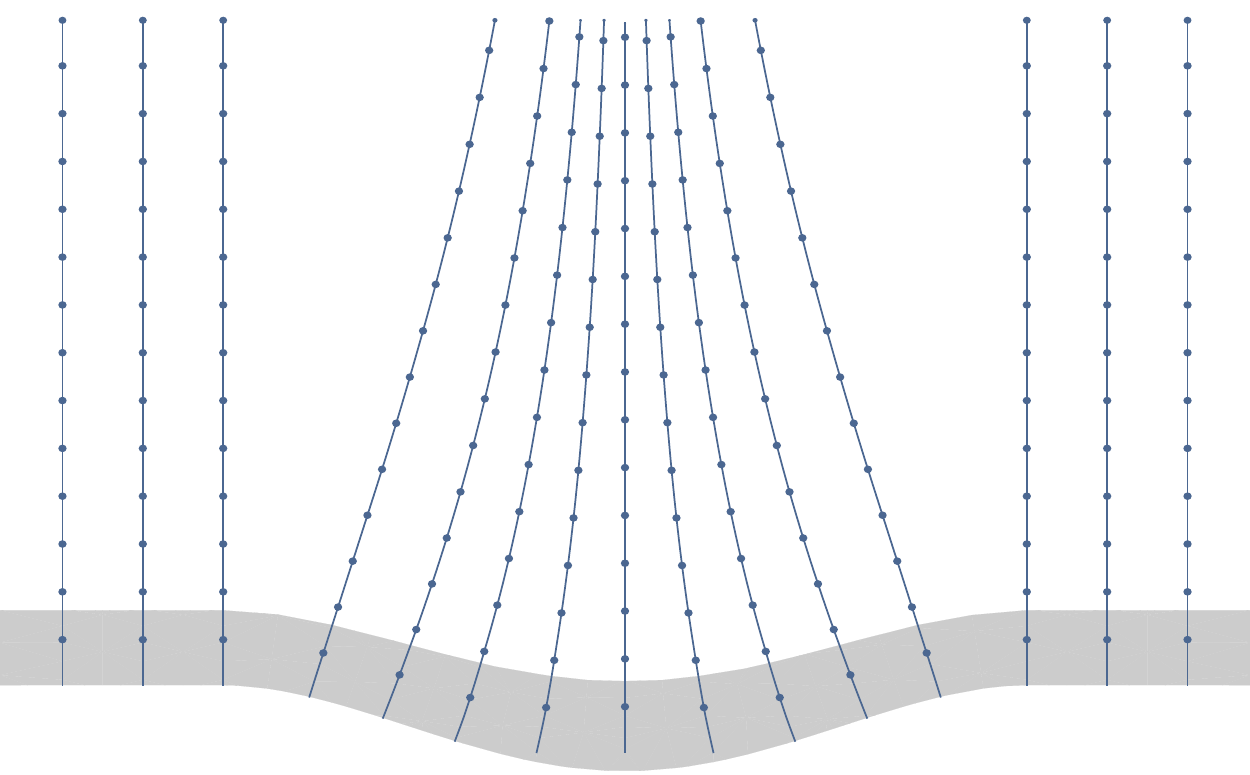}
\end{center}
\caption{
{Dust trajectories above concave areas.} 
Homogeneously seeded dust particles are entrenched and accelerated by outflowing gas within the outer mantle. Concavely shaped surface areas lead to the collimation of dust.
}
\end{figure}

\begin{figure}[h]
\begin{center}
\includegraphics[width=0.7\textwidth]{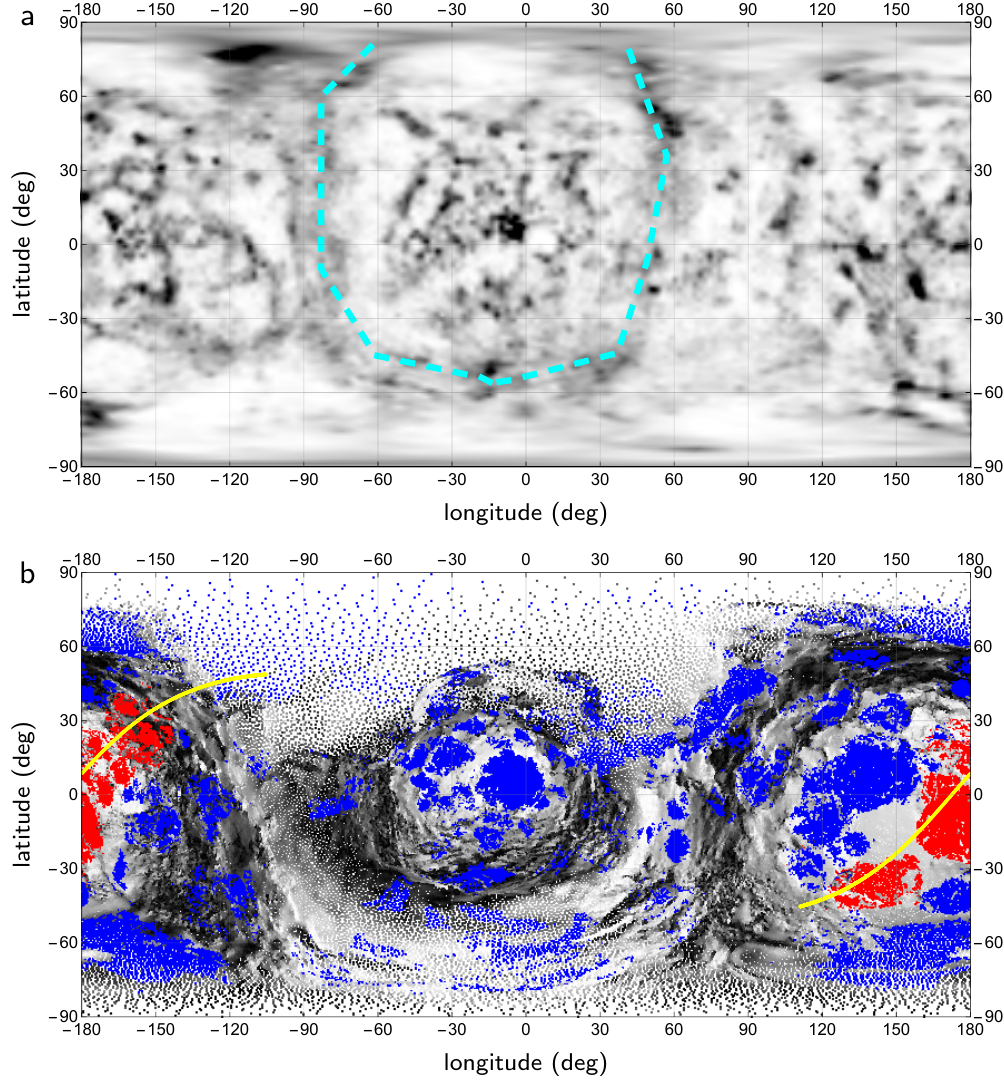}
\end{center}
\caption{
{Dust density around 67P originating from a homogeneous surface emission.}  
{\sf a} Equirectangular projection of the dust density at 3~km distance from the center of the nucleus. Darker spots indicate higher dust densities within primary jets. The dashed line marks a secondary dust belt encircling the neck area.
{\sf b} Map of the primary jets traced back to the surface. Blue and red surface areas contribute to primary jets. The yellow track indicates the line of sight from Rosetta on 12 April 2015, 12:12 along the highest observed intensity.
Surface areas connected to the 13 primary jets intersecting the line of sight are marked in red.
}
\end{figure}

\begin{figure}[h]
\begin{center}
\includegraphics[width=0.8\textwidth]{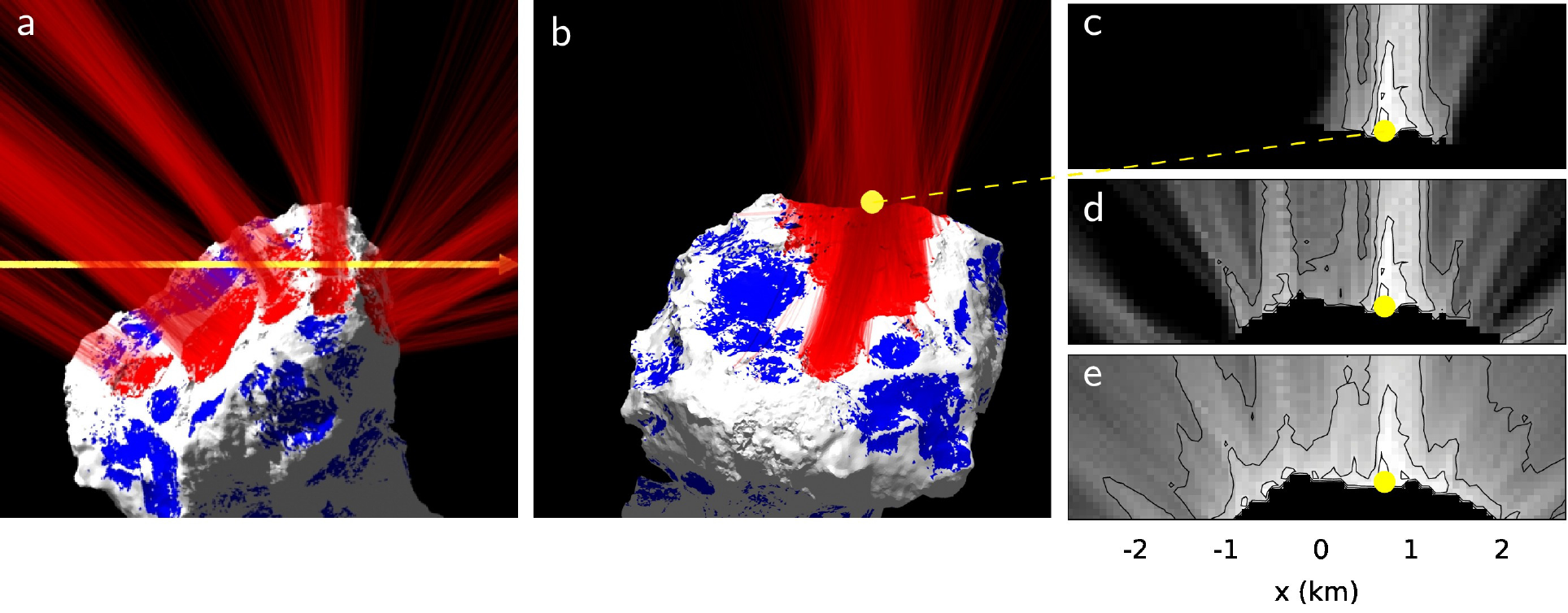}
\end{center}
\caption{
{Surface areas connected to primary jet formation on 12 April 2015 12:12.}
{\sf a} Rotated view, showing the intersection of the line of sight from Rosetta/OSIRIS (yellow) with the dust trajectories.
{\sf b} View along the line of sight with overlays of primary jets originating from different surface areas.
{\sf c} Dust column-density corresponding to {\sf b}, with contributions of only primary jets connected to red regions in Fig.~2{\sf b} and {\sf a,b}. 
{\sf d} Dust column-density with added contributions of all blue regions.
{\sf e} Dust column-density with dust arising from illuminated surface areas.
}
\end{figure}

\begin{figure}[h]
\begin{center}
\includegraphics[width=0.7\textwidth]{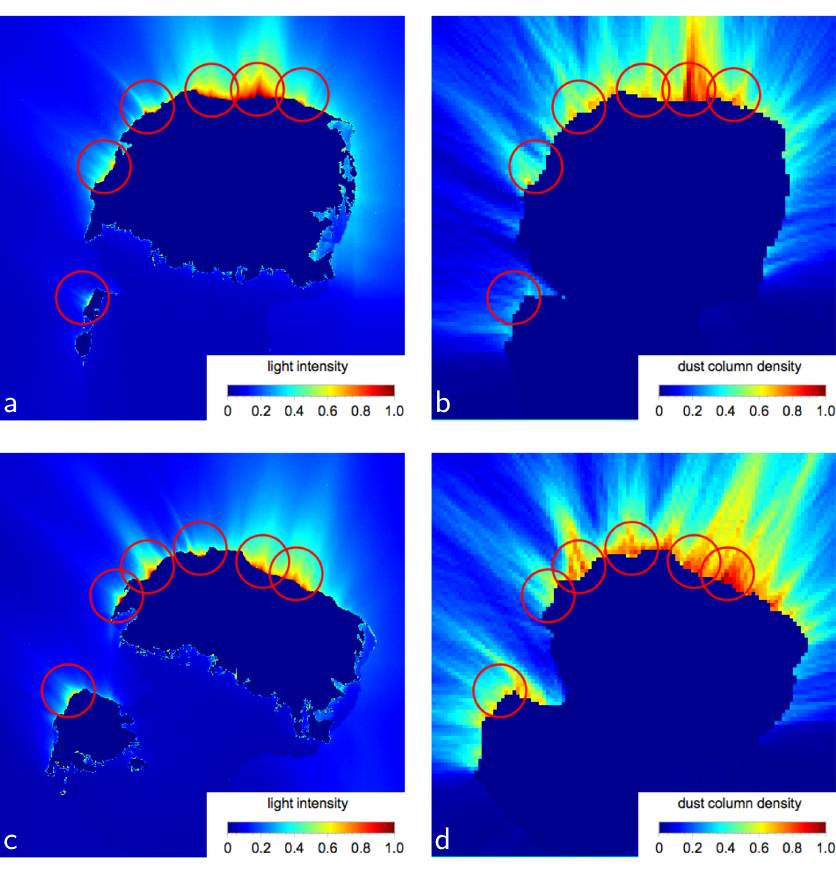}
\end{center}
\caption{{Comparison of Rosetta observations with the dust model.} OSIRIS NAC observation on (a) 12 April 2015 at 12:12, (c) 13:12  and calculated dust column densities (b,d) {not intersecting the nucleus, with the highest column density set to 1.0.
OSIRIS data has been linearly stretched and resulting overexposed pixels on the cometary surface have been set to zero (a,c).}
The circles mark the highest photographed intensities and matching regions of predicted increased dust column density. 
Credit for (a,c): ESA/Rosetta/MPS for OSIRIS Team MPS/UPD/LAM/IAA/SSO/INTA/UPM/DASP/IDA.
}
\end{figure}

\begin{figure}[h]
\begin{center}
\includegraphics[width=1.0\textwidth]{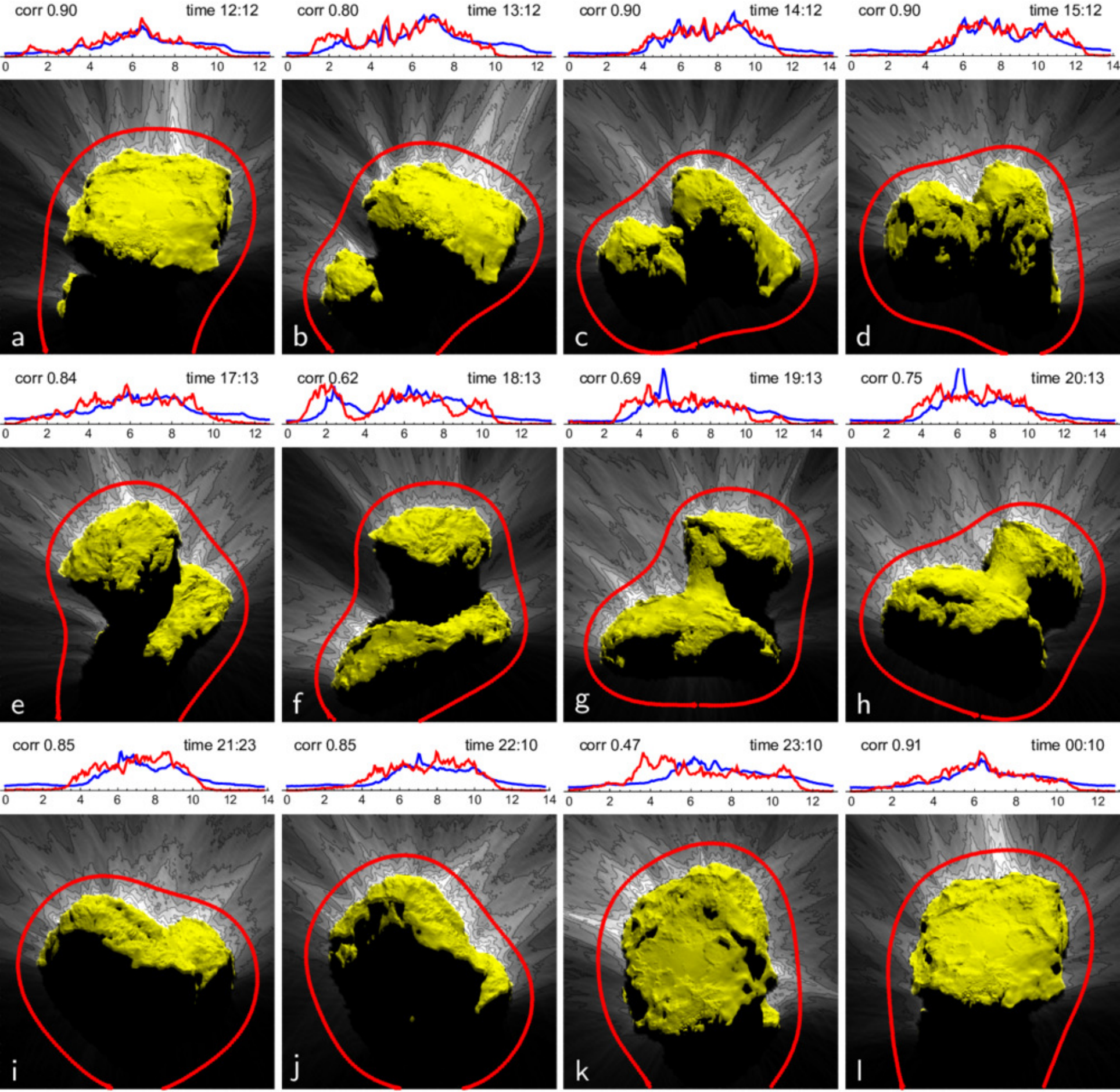}
\end{center}
\caption{
{Dust column-densities over a diurnal rotation period.} 
{Linear grayscale representation of the dust column-density resulting from the homogeneous surface activity model across the entire sunlit nucleus, normalized to the highest density within each panel.}
The yellow inset shows the illumination condition of the nucleus, the direction towards the sun is pointing up.
The viewpoints match  Rosetta's flight path and observations during 12 hours, starting on 12~April 2015, 12:12. 
The predicted dust column-density in the top panels (red) is compared to the image intensity recorded by Rosetta OSIRIS/NAC (blue) along the marked annulus around the nucleus.
The correlation coefficient (corr) of predicted dust-column density and observed light intensity is given in the insets.
}
\end{figure}

\end{document}